\documentclass{aa}  

\usepackage{graphicx}
\usepackage{txfonts}
\usepackage[colorlinks=true,allcolors=blue]{hyperref}
\newcommand{\fesc}{$f_{\rm{esc}}$\xspace}
\newcommand{\uvbeta}{$\beta_{\rm UV}$\xspace}

\begin{document} 

\title{Breaking Through the Cosmic Fog: JWST/NIRSpec Constraints on Ionizing Photon Escape in Reionization-Era Galaxies
}
\titlerunning{$f_{\rm esc}$ in the EoR with NIRSpec}

\author{Emma Giovinazzo\inst{1}\thanks{E-mail: emma.giovinazzo@unige.ch}
\and
Pascal A. Oesch\inst{1, 2, 3}
\and
Andrea Weibel\inst{1}
\and
Romain A. Meyer\inst{1}
\and
Callum Witten\inst{1}
\and 
Aniket Bhagwat\inst{4}
\and 
Gabriel Brammer\inst{2, 3}
\and 
John Chisholm\inst{5}
\and
Anna de Graaff\inst{6}
\and
Rashmi Gottumukkala\inst{2, 3}
\and 
Michelle Jecmen\inst{5}
\and
Harley Katz\inst{7}
\and 
Joel Leja \inst{8, 9, 10}
\and
Rui Marques-Chaves\inst{1}
\and
Michael Maseda\inst{11}
\and 
Irene Shivaei\inst{12}
\and
Maxime Trebitsch\inst{13}
\and
Anne Verhamme\inst{1}} 
\institute{
Department of Astronomy, University of Geneva, Chemin Pegasi 51, 1290 Versoix, Switzerland
\and 
Cosmic Dawn Center (DAWN), Niels Bohr Institute, University of Copenhagen, Jagtvej 128, K\o benhavn N, DK-2200, Denmark
\and
Niels Bohr Institute, University of Copenhagen, Jagtvej 128, Copenhagen, Denmark
\and
Max Planck Institut für Astrophysik, Karl Schwarzschild Straße 1, D-85741 Garching, Germany
\and 
Department of Astronomy, The University of Texas at Austin, Austin, TX 78712, USA
\and 
Max-Planck-Institut f\"ur Astronomie, K\"onigstuhl 17, D-69117 Heidelberg, Germany
\and
Department of Astronomy \& Astrophysics, University of Chicago, 5640 S Ellis Avenue, Chicago, IL 60637, USA
\and 
 Department of Astronomy \& Astrophysics, The Pennsylvania State University, University Park, PA 16802, USA
\and 
Institute for Computational \& Data Sciences, The Pennsylvania State University, University Park, PA 16802, USA
\and 
Institute for Gravitation and the Cosmos, The Pennsylvania State University, University Park, PA 16802, USA
\and
Department of Astronomy, University of Wisconsin-Madison, 475 N. Charter St., Madison, WI 53706 USA
\and 
Centro de Astrobiología (CAB), CSIC-INTA, Ctra. de Ajalvir km4, Torrejón de Ardoz, E-28850, Madrid, Spain
\and
LUX, Observatoire de Paris, Université PSL, Sorbonne Université, CNRS, 75014 Paris, France
}

   \date{Received XXX; accepted XXX}



\abstract
{{}
{The escape fraction of Lyman continuum photons ($f_{\rm esc}(\rm LyC)$) is the last key unknown in our understanding of cosmic reionization. Directly estimating the escape fraction (\fesc) of ionizing photons in the epoch of reionization (EoR) is impossible, due to the opacity of the intergalactic medium (IGM). However, a high \fesc leaves clear imprints in the spectrum of a galaxy, due to reduced nebular line and continuum emission, which also leads to bluer UV continuum slopes (\uvbeta). Here, we exploit the large archive of deep JWST/NIRSpec spectra from the DAWN JWST Archive to analyze over 1'400 galaxies at $5<z_\mathrm{spec}<10$ and constrain their \fesc based on SED fitting enhanced with a picket fence model.}
{We identify 71 high-confidence sources with significant \fesc based on Bayes factor analysis strongly favouring \fesc> 0 over \fesc= 0 solutions. We compare the characteristics of this high-escape subset against both the parent sample and established diagnostics including \uvbeta slope, O32, and SFR surface density ($\Sigma_{\rm SFR}$).}
{For the overall sample, we find that most sources have a low escape fraction (<1$\%$), however, a small subset of sources seems to emit a large number of their ionizing photons into the IGM, such that the average \fesc is found to be $\sim$10\%, as needed for galaxies to drive reionization.} 
{Although uncertainties remain regarding recent burstiness and the intrinsic stellar ionizing photon output at low metallicities, our results demonstrate the unique capability of JWST/NIRSpec to identify individual LyC leakers, measure average \fesc and thus constrain the drivers of cosmic reionization.}}


\maketitle



\section{Introduction}\label{ch:introduction}

The epoch of reionization (EoR) marks the last phase transition of the Universe, when the intergalactic medium (IGM) went from completely neutral to completely ionized \citep[for some recent reviews]{Becker_15, Dayal_18,Robertson_22, Fan_23, Stark_25}. This process starts with the formation of the first stars \citep{Barkana_Loeb_01} and does not conclude until $z \sim 5.3$, as suggested by the scatter in opacity measured from the Lyman-$\alpha$ and Lyman-$\beta$ absorption in quasar spectra  \citep[e.g.,][]{Eilers_18, Kulkarni_19, Yang_20, Bosman_22}.
Reionization is likely driven by star forming galaxies \citep[e.g.][]{Bouwens_15, Robertson_15, Naidu_20, Atek_24} which can leak Lyman Continuum (LyC) photons ($\lambda_{\rm rest}<912$ \AA) into the IGM. In this framework, the reionization process is very patchy  \citep[e.g.,][]{DAloisio_15, Bosman_18, Eilers_18, Yang_20, Bosman_22, Jamieson_24, Meyer_25}, with bubbles forming around the objects that leak the most ionizing photons, which then expand and overlap, until the entire Universe is ionized. 
AGNs have also been suggested as possible contributors to reionization \citep{Giallongo_15, Madau_15, Grazian_24, Madau_24}, although other studies of the AGN luminosity function in the EoR have shown that their contribution is likely minor, compared to that of galaxies \citep{Mitra_18, Hassan_18, Matsuoka_23}.

To determine which sources are responsible for the bulk of reionization, we need to determine $\dot n_{\rm ion}$, the number of ionizing photons that reach the IGM per unit time and volume \citep{Madau_Dickinson_14,Robertson_22}. If we can rely on the UV light to trace the bulk of star-formation in galaxies, this quantity, to first order, can be expressed as \begin{equation}\label{eq:ndotion}
    \dot n_{\rm ion} = f_{\rm esc}\xi_{\rm ion}\rho_{\rm UV}
\end{equation}
where $\rho_{\rm UV}$ is the UV luminosity density (erg s$^{-1}$ Hz$^{-1}$ Mpc$^{-3}$), calculated from the integral of the UV luminosity function, $\xi_{\rm ion}$ is the ionizing photon production efficiency (Hz erg$^{-1}$), which measures the number of ionizing photons created per 1500 \AA\ UV luminosity density and \fesc is the escape fraction of Lyman continuum photons, so the number of photons that escape the interstellar and circumgalactic media and reach the IGM. 

Building on the results from the Hubble Space Telescope (HST), the \textit{James Webb Space Telescope} (JWST) has allowed us to measure both  $\rho_{\rm UV}$ \citep{Bouwens_23, Donnan_24,Harikane_25, Whitler_25} and $\xi_{\rm ion}$ \citep{Simmonds_24a, Simmonds_24b,Llerena_24, Pahl_25} during reionization. Therefore, one of the last major unknowns to understand reionization is \fesc, making this a key quantity to identify the drivers of reionization. 

Predictions from models \citep{Bouwens_15,Robertson_13,Robertson_15, Finkelstein_19, Naidu_20} indicate an average \fesc$\sim5-20\%$ over cosmic time for Equation \ref{eq:ndotion} to work with globally-averaged quantities, although this does not take into account obscured star formation \citep{Simmonds_24}. Simulations that account for this source of ionizing photons predict lower \fesc values, such as $<5\%$ in \textsc{Sphinx} \citep{Rosdahl_22} and $\sim5\% - 10\%$ at redshift $6<z<10$ for \textsc{Thesan} \citep{Yeh_23} and \textsc{Obelisk} \citep{Trebitsch_21}.

As important as \fesc is, placing constraints on its value at $z>6$ in the EoR is highly challenging, due to the still partly neutral IGM. This is completely opaque to LyC photons and prevents the detection of Lyman continuum photons at $z>4.5$, making \textit{direct} observations of the escaped LyC flux and estimates of \fesc in the EoR impossible \citep{Inoue_14}. 

Until now, we have heavily relied on proxies and indirect indicators, calibrated at low redshifts, where the Lyman continuum leakage can be directly measured, to estimate the \fesc of epoch of reionization galaxies. Many studies have attempted to link various properties of known low-redshift Lyman continuum leakers to their \fesc. 
The most comprehensive of these studies is the Low-Redshift Lyman Continuum Survey \citep[LzLCS,][]{Flury_22a, Flury_22b} which observed 66 low-z ($z=0.2-0.4$) galaxies, of which 12 are LyC emitters with \fesc$>5\%$, and tested various indirect diagnostics. Other studies have tested the feasibility of the [\ion{Mg}{II}] line \citep[e.g.,][]{Chisholm_20, Xu_22}, the \ion{C}{IV} line \citep[e.g.,][]{Saxena_22, Schaerer_22}, the [\ion{O}{III}]$\lambda5007$/[\ion{O}{II}]$\lambda3727$ line ratio \citep[$O_{32}$; e.g.,][]{Jaskot_13,Nakajima_Ouchi_14, Izotov_18b, Paalvast_18, Tang_21}, the ultraviolet (UV) $\beta$ slope \citep[\uvbeta; e.g.,][]{Chisholm_22, Flury_22b}, star formation rate surface density \citep[$\Sigma_{\rm SFR}$;][]{Naidu_20}, various properties of the Lyman-$\alpha$ line such as the double-peak separation \citep[$v_{\rm sep}$;][]{Verhamme_17,Izotov_18b, Naidu_Matthee_22} or even multiple indicators together \citep[e.g.,][]{Mascia_23,Mascia_24,Jaskot_24a, Jaskot_24b}. However, many of these relations present large amounts of scatter and, although they have been used to estimate \fesc at high redshift \citep[e.g.][]{Navarro_Carrera_24}, it is not obvious that relations calibrated on sub-samples of galaxies at low redshifts, with often unclear selection functions, will necessarily hold for the average galaxy in the EoR. Indeed, \cite{Pahl_24} find that this is not the case, at least when using  the Lyman-$\alpha$ line shape as a proxy for Lyman continuum escape. They find different relations between \fesc and the Lyman-$\alpha$ line shape at $z\sim0.3$ and $z\sim3$. Moreover, \cite{Witten_23} have found that the fraction of neutral gas in the IGM has a strong impact on the red peak of Lyman-$\alpha$, which would affect the shape of the line through cosmic time and \cite{Giovinazzo_24} have found the Lyman-$\alpha$ line alone is not sufficient to estimate LyC \fesc in the EoR. Overall, a big caveat in the use of low-redshift analogues are the intrinsic differences between the low and high redshift Universe, due to differences in environment, leading to fewer mergers and inflow of less pristine gas at later cosmic times, all of which can have an impact on the escape fraction. 
It is therefore evident that other methods for estimating \fesc in the EoR are needed.

Thanks to JWST and its NIRSpec instrument \citep{Jakobsen_22} we now have deep rest-UV and rest-optical spectra of galaxies directly in the EoR. This gives us access to all the information encoded in the spectra, including information on \fesc. Indeed, we expect the higher escape of ionizing photons to correlate with reduced nebular line and continuum emission, given the conservation of ionizing photons. Since nebular continuum emission and the presence of dust make the \uvbeta slope redder we can expect steep \uvbeta slopes and weak nebular line emission (e.g., H$\beta$) to link with high \fesc \citep{Zackrisson_13,Zackrisson_17, Chisholm_22, MarquesChaves_22,Topping_22, MarquesChaves_24, Yanagisawa_24} . 

In this work, we follow this approach, exploiting the large public archive of NIRSpec/PRISM spectra from a range of public JWST programs to analyze 1428 spectra of galaxies in the EoR and perform Spectral Energy Distribution (SED) fitting with a picket-fence model directly on the spectra to estimate their \fesc. For similar works see also \citet{Papovich_25}.

In Section~\ref{ch:Data}, we present the dataset that we use to perform this analysis. In Section~\ref{ch:Methods} we present the picket-fence model, the SED fitting code we use, the selection of the high-confidence sample, and the line and sizes measurements. Then, in Section~\ref{ch:Validation} we validate our method with a recovery simulation. In Section~\ref{ch:Results} we present the main results and compare to other methods in Section~\ref{ch:Comparison}. We discuss possible degeneracies and selection effects in Section~\ref{ch:discussion} and finally conclude in Section~\ref{ch:Conclusion}.

Throughout this work, we assume flat $\Lambda$CDM cosmology with $H_0=70\,\textrm{km}\,\textrm{s}^{-1}\,\textrm{Mpc}^{-1}$, $\Omega_m=0.3$, and $\Omega_\Lambda=0.7$. Magnitudes are given in the AB system \citep{Oke_Gunn_83}. 

\section{Data}\label{ch:Data}

\subsection{NIRSpec prism spectroscopy}\label{ch:spectroscopy}

The data in this work is part of the DAWN \textit{JWST} Archive (DJA\footnote{\url{https://dawn-cph.github.io/dja/}}) \citep{Heintz_25}, an online repository containing spectroscopic data from public \textit{JWST} programs, all uniformly extracted and reduced with the same pipeline, which makes use of \texttt{grizli}\footnote{\url{https://github.com/gbrammer/grizli}} and \texttt{MSAexp}\footnote{\url{https://github.com/gbrammer/msaexp}}. Further details on the data reduction and processing can be found in \citet{deGraaff_24,Heintz_25,Valentino25,Pollock25}. We performed our database query on the 7th of February 2025 and include all sources with available PRISM/CLEAR spectra at $5<z<10$ with grade$=$3 (i.e., with a robust redshift measurement based on visual inspections) and with available photometry in at least one JWST NIRCam filter, as described in Section~\ref{sec:photometry}, which is needed to estimate sizes and slitloss corrections. 

\begin{figure}
    \centering
    \includegraphics[width=\linewidth]{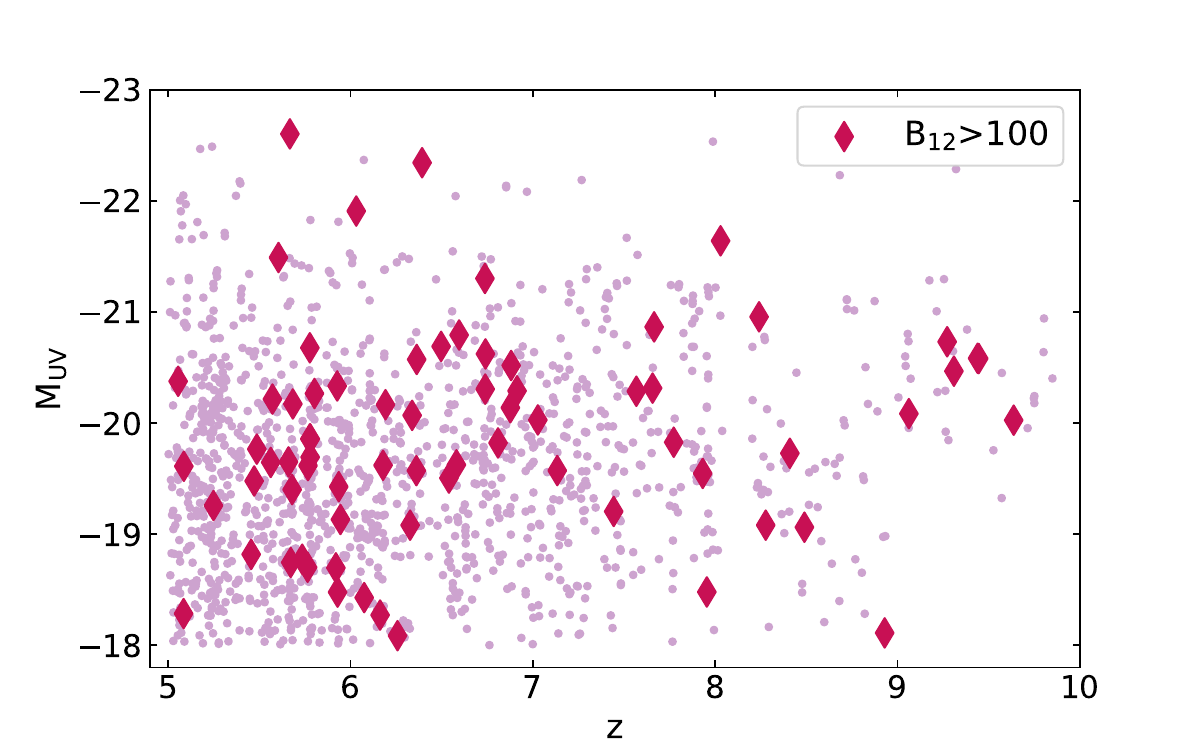}
    \caption{Distribution of M$_{\rm UV}$ as a function of redshift for the whole sample. The high confidence sample (see Section~\ref{sec:high_conf}) is highlighted with diamond markers, while the parent sample of spectroscopic redshifts is shown as dots. No significant trends or biases in M$_{\rm UV}$ are seen.}
    \label{fig:Muv_z}
\end{figure}

The spectra stem from various public surveys, including RUBIES (GO 4233, \citealt{deGraaff_24})  JADES (GTO-1180, 1181, 1210, 1286, 1287; GO-3215 \citealt{Eisenstein_23a}), GO-3215 (\citealt{Eisenstein_23b}) GTO WIDE (GTO-1211, 1212, 1213, 1214, 1215; \citealt{Maseda_24}), CEERS (ERS-1345; \citealt{Finkelstein_23}), UNCOVER (GO-2561, \citealt{Bezanson_24}), CAPERS (GO-6368; PI: Dickinson), GO-1433 \citep{Hsiao_24}, GO-2198 \citep{Barrufet_25}, GO-2565 \citep{Nanayakkara_24}, DD 2750 \citep{ArrabalHaro_24}, GO-3073 \citep{Castellano_24}, DD-2756 (PI: Chen), DD-6541 (PI: Egami), DD-6585 (PI: Coulter), GO-4106 (PI: Nelson). The number of sources corresponding to each program can be found in Table~\ref{tab:DJA}. The redshift constraints that we apply to our sample allows us to select sources in the epoch of reionization that also have coverage of the H$\beta$ line, which is bright enough to be detected in most galaxies and is also visible at high redshifts. We apply a UV magnitude cut at $\rm M_{\rm UV} = -18$. The distribution of $\rm M_{\rm UV}$ as a function of redshift is shown in Figure~\ref{fig:Muv_z}, which does not show any significant evolution or biases in UV magnitudes at different redshift. Lastly, we remove 64 broad-line LRDs reported in \cite{Kocevski_24}. We do not remove any other AGNs as they are unlikely to have a significant impact on the identification of leakers in our framework where we expect high \fesc sources to have very faint lines. In total, our sample consists of 1'428 galaxies at $5<z_{\rm spec}<10$. While this wide selection gives us access to a large quantity of data, it also leads to an effectively unknown selection function with varying exposure times. The possible impact of this will be addressed in Section~\ref{ch:discussion}.

\begin{table}
    \centering
    \caption{Overview of the sources included in our analysis.}
    \label{tab:DJA}
    \begin{tabular}{c|c|c}
    Program-ID & Field & $N_{\rm sources}$ \\ \hline
        GTO 1210 & GOODS-S & 30  \\
        GTO 1211 & GOODS-N & 26  \\
        GTO 1212 & GOODS-S & 11  \\
        GTO 1213 & EGS & 8  \\
        GTO 1214 & COSMOS & 10 \\
        GTO 1215 & UDS & 13  \\
        ERS 1345 & EGS & 77  \\
        GTO 1180 & GOODS-S & 96\\
        GTO 1181 & GOODS-N & 157   \\
        GTO 1286 & GOODS-S & 129 \\
        GTO 1287 & GOODS-S & 15  \\
        GO 1433 & MACS 0647 & 9  \\
        GO 2198 & GOODS-S & 17   \\
        GO 2561 & Abell 2744 & 117 \\
        GO 2565 & COSMOS/EGS/UDS & 10  \\
        DD 2750 & EGS & 16  \\
        DD 2756 & Abell 2744 & 11  \\
        GO 4106 & EGS & 35 \\ 
        GO 4233 & EGS/UDS & 452  \\
        GO 3073 & Abell 2744 & 34  \\
        GO 3215 & GOODS-S & 18 \\
        DD 6541 & GOODS-S & 17 \\
        GO 6368 & UDS & 89  \\
        DD 6585 & COSMOS & 31  \\ \hline
        
        Total & & 1428 \\
    \end{tabular}

\end{table}

\subsection{Photometry}\label{sec:photometry}

Based on the JWST and ancillary HST imaging available on the DJA, we derive photometric catalogs following \cite{Weibel_24}. We use an inverse-variance weighted stack of the long wavelength filters F277W, F356W, and F444W as the detection image, and measure fluxes in circular apertures of 0.16\arcsec\ radius in point spread function (PSF) matched images. These aperture fluxes are scaled to total based on the flux measured through Kron ellipses in the detection image and an additional correction to account for flux in the wings of the PSF.

The photometric data are needed to correct flux sensitive quantities such as UV magnitudes and masses, as in many cases the slit is positioned such that only part of the galaxies is observed. Using the photometry helps us estimate slitloss corrections. We perform these corrections by creating synthetic photometry from the spectra and by scaling this synthetic photometry to the real photometry, using a wavelength-independent scaling. The typical rescaling factors hover around 1, with a median of 1.1, indicating that the \texttt{msaexp} corrections are accurate and that most of our sources are compact, as expected at high redshift. We use this uniform scaling to correct the UV magnitudes and the masses estimated from the spectra. 

We also use the photometry to fit for galaxy sizes. The fitting process is described in detail in Section~\ref{ch:size_fits}.


\section{Methods}\label{ch:Methods}

\subsection{Picket-fence Model}

\begin{figure}
    \centering
    \includegraphics[width=\columnwidth]{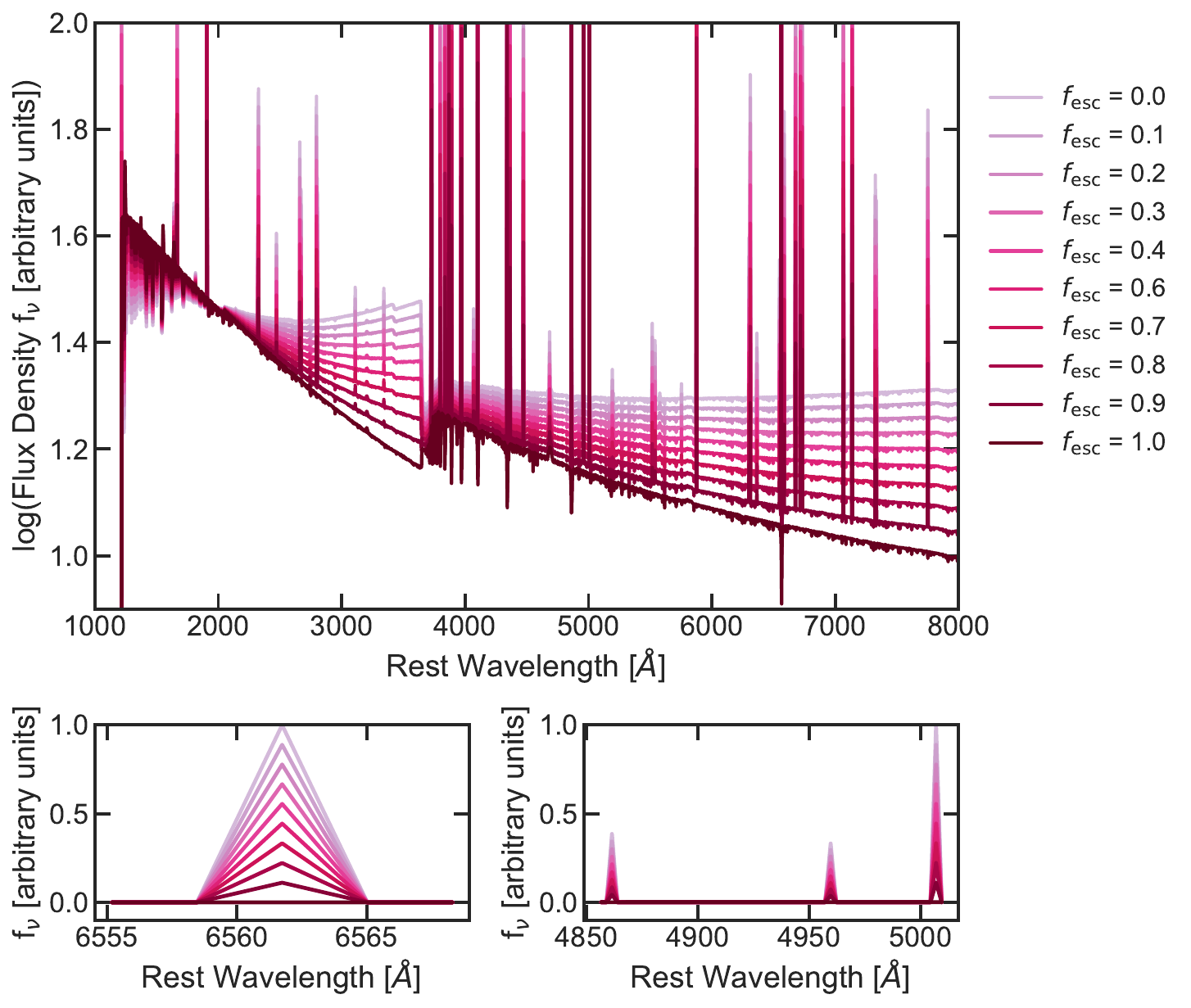}
    \caption{ \textbf{Top}: Spectrum of a model galaxy with various \fesc. The model galaxy is at  $z=6$, has a mass of $\rm M_{*} = 10^9 M_{\odot}$, metallicity $Z= 0.05 Z_{\odot}$, ionization parameter $\rm logU = -2$, constant star formation  switched on at 5 Myr and dust modelled with the Calzetti \citep{Calzetti_00} dust curve with $\rm A_{\rm v} = 0.2$.
    As \fesc increases the \uvbeta slope becomes steeper due to reduced nebular continuum emission, and the emission lines become weaker. The continuum emission is also affected, as its nebular component scales with \fesc. The full spectrum thus contains information on the escape fractions, which we exploit to constrain the \fesc of galaxies with NIRSpec spectra.
    \textbf{Bottom left}: Zoom in on H$\alpha$. \textbf{Bottom right}: Zoom in on H$\beta$+[\ion{O}{iii}]}
    \label{fig:model_galaxy}
\end{figure}

To estimate \fesc with \texttt{bagpipes} we implement a "picket-fence" \citep{Heckman_01,Zackrisson_13} model. With this model we assume that stars are partially covered by an optically thick ISM, with some channels of low column density (and no dust) which allow for ionizing photons to escape. Effectively, this means that only a fraction of a source is covered by optically thick gas, so that \fesc is estimated as $f_{\rm esc} = 1- C_{f}$ where $C_f$ is the covering fraction. In the SED fitting, this means that the output spectra will be a linear combination of $f_{\rm esc} = 0$ models in the fraction that is covered by gas and $f_{\rm esc} = 1$ and $A_{V} = 0$ models in the free channels. While this model has been found to not exactly reproduce the \fesc from radiative transfer in simulations, it is in reasonable agreement \citep{Mauerhofer_21} and has been used before at redshift $3<z<5$ to connect UV absorption features to \fesc \citep{Saldana_Lopez_23}.

The effects of the picket-fence model on a spectrum are shown in Figure~\ref{fig:model_galaxy}. Note that we would expect similar traces of \fesc on the spectra from a density-bounded model, at least in a dust free scenario \citep{Zackrisson_13}. The galaxy modelled in Figure~\ref{fig:model_galaxy} is at redshift $z=6$, has a stellar mass of $\rm M_{*} = 10^9 \rm M_{\odot}$, a metallicity of $\rm Z = 0.05 \rm Z_{\odot}$, an ionization parameter $\rm logU = -2$, a dust attenuation of $\rm A_{\rm V} = 0.2$, modelled with a Calzetti \citep{Calzetti_00} dust model and a constant star formation history (SFH) which is switched on at 5 Myr. The only parameter that we modify between the various models is \fesc to highlight the effect of this quantity on a mock galaxy. As \fesc increases, there is a clear steepening of the \uvbeta\ slope and weakening of all the emission lines, highlighted in the bottom panels -- one for $\rm H\alpha$ (left) and one for $\rm H\beta$+[\ion{O}{III}] (right). There is also an effect on the continuum in the optical part of the spectrum, which is reduced due to an overall reduction or absence of the nebular continuum contribution. 

It is possible for the \uvbeta slope to redden due to dust or an older stellar population, however here we are just highlighting the effects of \fesc on the spectrum. These effects should make it possible to estimate it at any redshift, as long as the \uvbeta slope and emission lines are covered in the spectrum, which they are in the EoR when observing with the PRISM configuration of NIRSpec. 

\subsection{SED fitting}
We implement a picket-fence model in the \texttt{bagpipes} SED fitting code\footnote{available at \url{https://github.com/pascaloesch/bagpipes-wfesc}} \citep{Carnall_18, Carnall_19, nautilus}. This is then applied directly to the spectra to estimate the physical properties of our sources, including \fesc. Our custom version of \texttt{bagpipes} also employs updated CLOUDY grids \citep[][]{Cloudy17} that were run without internal dust (i.e., with \texttt{grains ISM} turned off). This solves an issue with the emission line normalizations in the original \texttt{bagpipes} grids.

The choice of free parameters, their ranges and priors are outlined in Table~\ref{tab:bagpipes_parameters}. The range for logU is based on the results of \citet{Reddy_23}. We also fit for the redshift, using a very narrow Gaussian prior, with a standard deviation of only 0.002, centred at the spectroscopic redshift $z_{\rm spec}$. We use the BPASS v2.2.1 stellar population models \citep{Stanway_Eldridge_18} with the default broken power law Initial Mass Function (IMF), with slopes of $\alpha_1 = -1.30$ for the mass range $0.1 - 0.5 \rm M_{\odot}$ and $\alpha_2=-2.35$ for the mass range $0.5 - 300 \rm M_{\odot}$, based on \cite{Kroupa_93}. To estimate the stellar and nebular attenuation we use the Calzetti dust law \citep{Calzetti_00}, allowing for a maximum $A_{V} = 0.5$ as we are mainly interested in blue sources. This choice does lead to poor fits for very dusty galaxies which are unlikely to be leakers and therefore will only be in the background sample. We model the SFH with the \texttt{continuity} prior \citep{Leja_19} in \texttt{bagpipes}, a non-parametric model, which allows for greater flexibility in the SFHs than parametric models would. Using this prior, we estimate the star formation rate in 7 bins, with 5, 10, 50, 100, 200, 400 and 800 Myr, unless the age of the Universe at the source's redshift was less than 800 Myr. In that case, the last bin will end at the age of the Universe. With the continuity prior, \texttt{bagpipes} fits for the $\Delta \rm SFR$ between adjacent bins which therefore adds a number of free parameters equal to the number of bins minus one. We use a Student-t prior with $\Delta \rm SFR \in [-3, 3]$ and adopt $\nu = 2$ and $\sigma = 0.3$ following \cite{Leja_19}. We opt for a logarithmic prior for \fesc as observations seem to suggest that most sources have little-to-no leakage \citep{Kreilgaard_24}, which is more consistent with a logarithmic than a constant distribution.  

When performing the fit, we choose to mask the spectrum below rest frame wavelengths of $\lambda_{\rm r}<1300$ \AA\ to avoid fitting the Lyman-$\alpha$ line as well as the IGM attenuation that can vary from source to source and may include damped Lyman-$\alpha$ (DLA) absorption profiles \citep{Heintz_24,Heintz_25b, Mason_25,Meyer_25}. 

\begin{table}
    \centering
    \caption{\texttt{Bagpipes} parameters used for the SED fitting.}
    \label{tab:bagpipes_parameters}
    \begin{tabular}{|c|c|c|}
       Parameter & Range & Prior \\ \hline
        log$_{10}(M/M_{\odot}$) & [5, 12] & linear  \\
        Metallicity (Z/Z$_{\odot}$) & [0.01, 0.5] & log \\
        logU & [-3.5, -1.5] & linear \\
        \fesc & [0.001, 1] & log \\
        A$_{\rm V}$ [mag] & [0, 0.5] & linear \\
        $\Delta \rm SFR$ & [-3, 3] & Student-t \\
    \end{tabular}
\end{table}

\subsection{High confidence sample selection}\label{sec:high_conf}

To identify which galaxies are more likely to be leakers we perform a second run with fixed \fesc = 0 on the entirety of our sample. We then compare the statistical evidence of the two different runs by calculating the Bayes factor. This will help us determine how likely a high \fesc solution is compared to a low \fesc solution and select galaxies for the high confidence sample. 
The Bayes factor, introduced by \cite{Jeffreys1939}, compares the evidence of two models with equal priors, and quantifies the support for one model over the other, given the data. It is defined as:
\begin{equation}
    B_{12} = \frac{p(y|M_{1})}{p(y|M_2)}
\end{equation}
where $M_1$ and $M_2$ are the models being tested and $y$ represents the data the models are being tested on. 
In our case, the models being tested are one with variable \fesc  and the $f_{\rm esc}=0$ model, given the priors and model used. Therefore, a high Bayes factor implies that $f_{\rm esc}>0$ is favored with respect to the $f_{\rm esc}=0$ solution. We compute the Bayes factors for the full spectra and use them to determine a high-confidence sample. To do this, we follow the table provided by \cite{Kass_95} and choose to consider Bayes factors greater than 100 as decisive evidence.
We therefore select the high-confidence sample as galaxies with Bayes factor $\rm B_{12}>100$.

With these criteria, we initially identify 74 high-confidence LyC leakers. We visually inspect all sources and remove 3 objects with poor fits. After the visual inspection, we are then left with 71 sources which we consider the high-confidence sample of LyC leakers. This sample will be identified in all the following scatter plots as the diamond markers. 

\begin{figure}
 \includegraphics[width=\columnwidth]{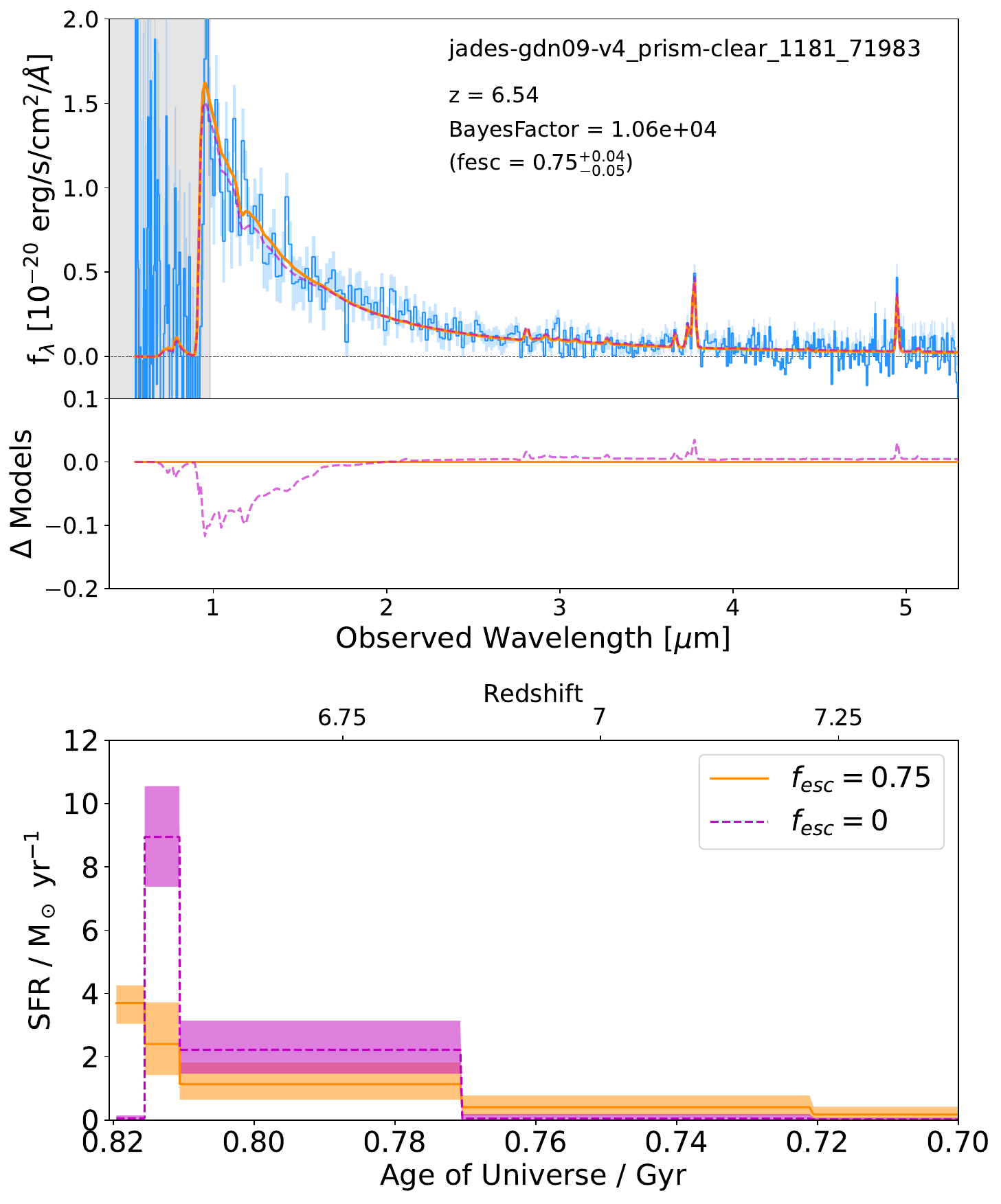}
    \caption{Example galaxy with a Bayes factor $>$ 100. \textbf{Top: } Here we show the observed spectrum (blue line) and the two models, one with high \fesc (orange solid line) and one with \fesc = 0 (magenta dashed line). The grey shaded region represents the masked region. A clear difference between the two models can be seen in the \uvbeta part of the spectrum, where the high \fesc solution fits the data much better than the other solution.
    \textbf{Middle: } Difference between the two models highlighting the difference in the \uvbeta slope and to some extent in the emission line strengths.
    \textbf{Bottom: } Comparison of the SFH for the two models. The models are extremely different, as reproducing the weak lines and steep \uvbeta slope without \fesc is only possible with a recent quenching of star formation. This indicates a degeneracy between SFH and \fesc, which is discussed more in Section~\ref{ch:discussion}. }
    \label{fig:rubies_spec}
\end{figure}

The spectrum of one of our high \fesc candidates, together with the two fits, is shown in the first panel of Figure~\ref{fig:rubies_spec}. For this source, the Bayes factor was calculated to be $1.06 \times 10^{4}$, which indicates the high \fesc solution to be a much better fit to the data. The most striking difference between the two models is in the \uvbeta slope, which the high \fesc model can fit well. The $f_{\rm esc}=0$ model cannot reproduce such a blue slope, resulting in a mismatch with the data. This is highlighted in the middle panel, where we show the difference between the two models.

In the same figure, we also show the SFH of the two models. The difference stems from the different mechanisms that need to be used by \texttt{bagpipes} to match the spectrum: high \fesc in one case and post starburst in the other. There is therefore some level of degeneracy between the estimation of \fesc and that of the SFH. This will be discussed in more detail in Section~\ref{ch:discussion}.

\subsection{Size measurements}\label{ch:size_fits}

One potential tracer of \fesc proposed by \cite{Sharma_16, Sharma_17} and \cite{ Naidu_20} is the star formation rate surface density. Hence, we also estimate the effective radii ($r_{\rm eff}$) of our sources. For this, we use imaging data from JWST/NIRCam in the F115W, F150W or F200W filters, depending on the redshift of the objects. As we are interested in the size in the UV, we select the filter which contains 1600 \AA\ in rest frame wavelength and if this wavelength is not covered by any filter, we use F115W as it is the bluest filter. Hence, we use F115W in the 5<$z$<7.3 range, F150W in the 7.3<$z$<9.3 range and F200W for $z>9.3$. The short wavelength filters are not always available so for some sources (107) it was impossible to calculate the sizes in the UV. 

The fitting is performed with \texttt{pysersic} \citep{Pasha2023}, using the variational inference method \texttt{svi-flow}. We use uniform priors for $r_{\rm eff}$ and $n$ with ranges of [0.25, 10] kpc and [0.65, 6], respectively. For the center pixels we use gaussian priors with the mean as the center of the cutout and a 1 pixel standard deviation. We mask all neighbouring sources and use the empirical PSF models described in \cite{Weibel_24}. All other priors were set with the \texttt{autoprior} function of \texttt{pysersic}.

\subsection{\uvbeta slope and line measurements}

The UV-continuum of galaxies can be characterised by a power law:
\begin{equation}
    f_{\lambda} \propto \lambda^{\beta}
\end{equation}
where $\beta$ is the UV-slope, \uvbeta.

We measure the UV slope from the best fit SED in the 1268<$\lambda$<2580 \AA\ range. We choose to exclude 10 windows, as described by \cite{Calzetti_94}, in order to avoid stellar and inter-stellar absorption features that can affect the shape of the continuum. This ensures that we are fitting the continuum and that our fit is not contaminated by the lines. We choose to measure the \uvbeta slope from the SED rather than from the spectra themselves to reduce scatter for UV-faint sources. 

Also from the best fit SED, we measure the equivalent width of H$\beta$ using the \texttt{indices} function within \texttt{bagpipes}. We take the line fluxes for [\ion{O}{III}] and [\ion{O}{II}] from the DJA\footnote{from \url{https://dawn-cph.github.io/dja/blog/2025/05/01/nirspec-merged-table-v4/}}, which have been extracted using \texttt{masexp}.

\section{Validation}\label{ch:Validation}

Before applying our method to all of the spectra, we perform some validation tests.
Since we are using data from very different programs, with different selections, it is necessary to understand how our method behaves at different SNR levels. We thus perform a recovery simulation to determine how well \texttt{bagpipes} can recover a known escape fraction from a model spectrum when noise is added. We perform this test on a model galaxy from \texttt{bagpipes}, with every model parameter fixed except for the escape fraction, which we vary, using 9 values. 
We then normalize the spectra to have results consistent with the SED fits on the NIRSpec spectra. To do this we calculate the median continuum flux at 5100 \AA$<\lambda_r<$6500 \AA\ for the overall sample and scale the each model spectra with a factor $f_{\rm norm} = \langle f_{\nu, \rm NIRSpec} \rangle / \langle f_{\nu, \rm model} \rangle$, where the median continuum flux for the model spectrum is calculated in the same wavelength range. 

We then pass the normalized model with the given escape fraction through the JWST exposure time calculator (ETC)\footnote{\url{https://jwst.etc.stsci.edu/}}, to degrade the \texttt{bagpipes} SED model to prism resolution. This also gives us the SNR curve, which we use to add the desired level of noise. 

We use four different SNR levels, 3, 5, 10 and 20. For each \fesc and SNR value we create three different instances of Gaussian noise that is added to the input spectra, thus giving us three realizations of spectra per \fesc and SNR value.
Each of the spectra is fit with \texttt{bagpipes} using the same set up as the one used for the real NIRSpec spectra. We also perform a \fesc= 0 run, to determine the Bayes factor. 
\begin{figure}
    \centering
    \includegraphics[width=0.85\columnwidth]{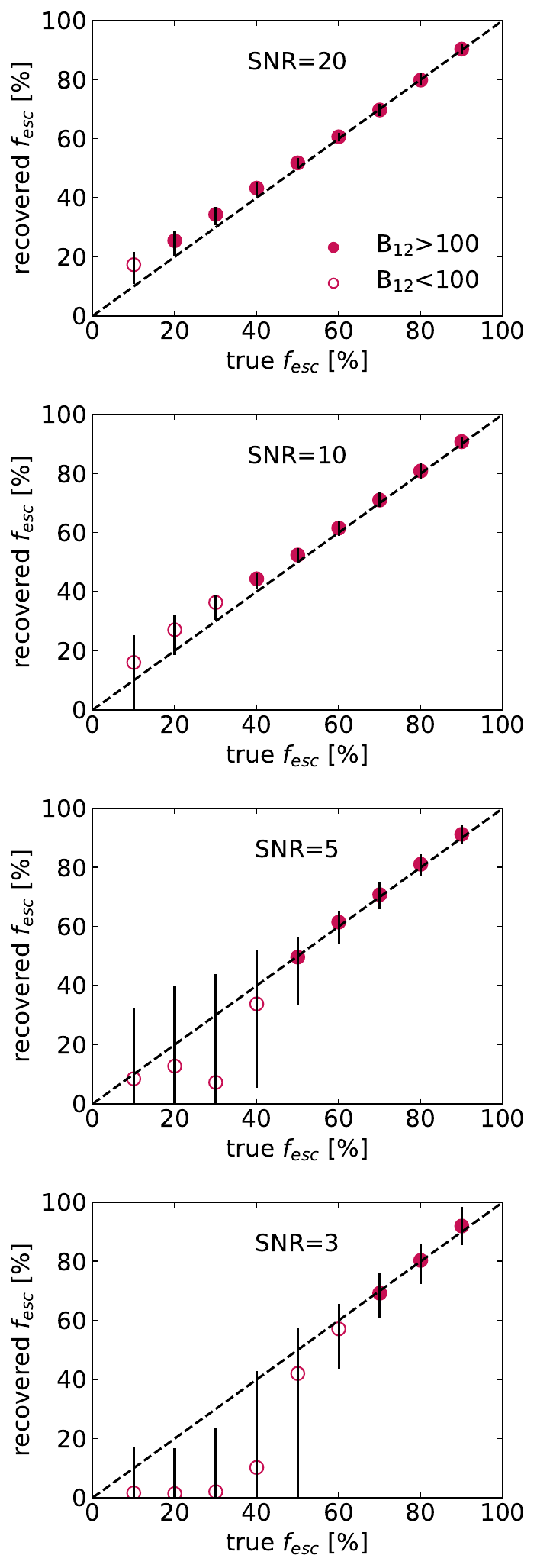}
    \caption{Results of the recovery simulation. The open markers indicate a Bayes factor of <100, while the filled markers indicate a Bayes factor of  >100. The black dashed line indicates the 1:1 line, so the true values. Our method can recover the high \fesc galaxies with a high confidence at all levels of signal to noise, while we can also recover lower \fesc values at high SNR levels.}
    \label{fig:recovery_simulation}
\end{figure}

The results of the recovery simulation are shown in Figure ~\ref{fig:recovery_simulation}, where each panel is a different SNR level. In all the panels, the filled points indicate that the Bayes factor is >100 while the empty points have a Bayes factor <100 and the dashed line indicates the 1:1 line, so the truth. At SNR=20, all the runs except for \fesc= 10\% have a high Bayes factor and while the high \fesc points are on the truth line, the low \fesc (<50\%) are slightly overestimated. In the SNR$\leq 10$ cases, the Bayes factors for the lowest \fesc values become <100, and the uncertainties increase. Overall, the results are consistent with the truth except when SNR=3, for all \fesc<30\%, where all the recovered \fesc are consistent with \fesc=0.
We can thus conclude that our method is robust and can safely recognize high \fesc sources even at low signal to noise levels.

\section{Results}\label{ch:Results}
Having validated our picket-fence model fitting on NIRSpec data, we will now present our results from the SED fitting on the sample described in Section \ref{ch:spectroscopy}. We will present the inferred \fesc and other properties of the sources. 

\subsection{\uvbeta vs EW$_{0}( \rm H\beta)$}
\begin{figure}
    \centering
    \includegraphics[width=1.05\columnwidth]{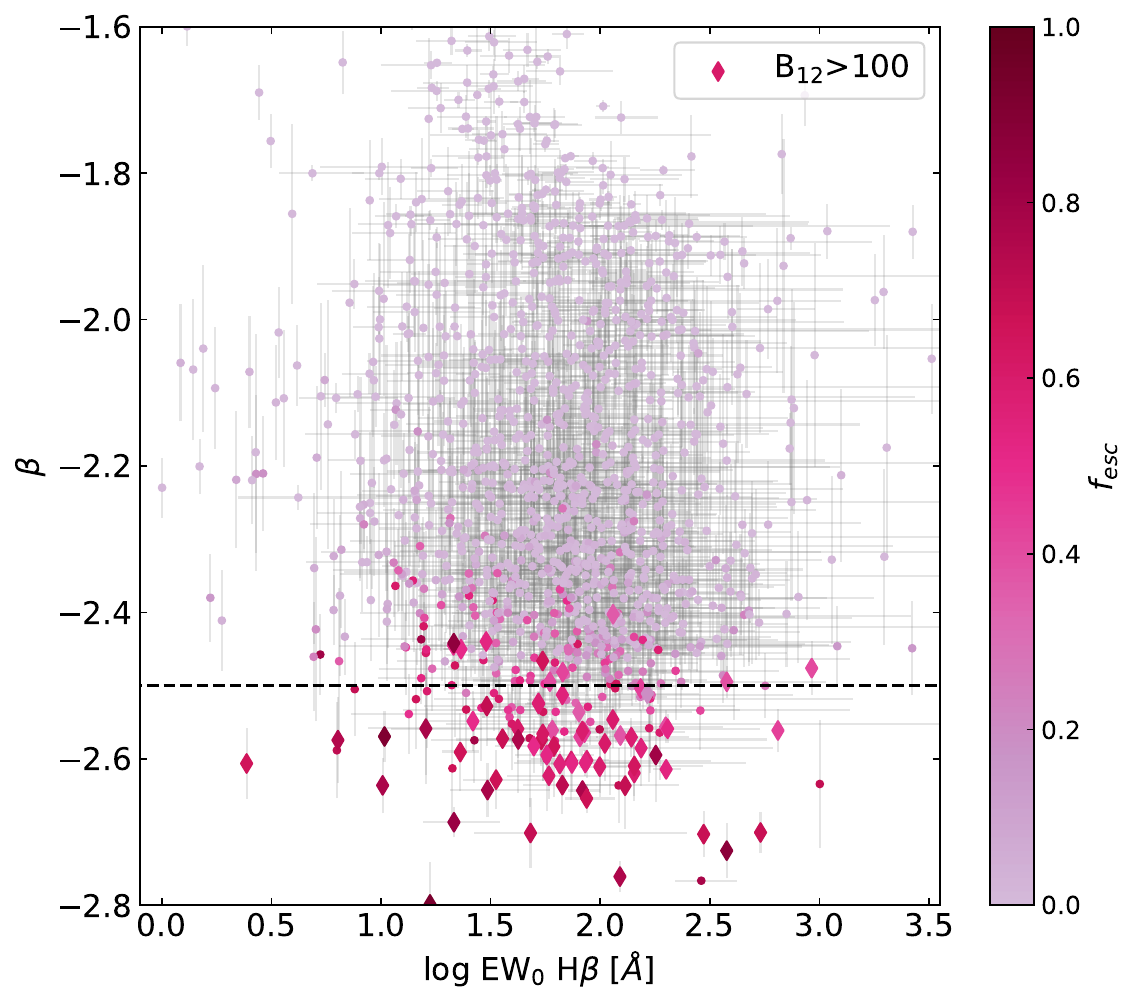}
    \caption{\uvbeta slope vs equivalent width of H$\beta$, color coded by \fesc. The parent sample is shown as circles and the high confidence sample is highlighted with diamond markers. All of the high confidence sources occur at low \uvbeta slope with most having an \uvbeta slope of <-2.5, but they span the entire EW range.}
    \label{fig:beta_vs_ew}
\end{figure}

We start by showing the relation between \uvbeta slopes, the rest frame equivalent width (EW${_{0}}$) of H$\beta$, both calculated from the best fit SED, and \fesc in Figure~\ref{fig:beta_vs_ew}. The \uvbeta - EW$_{0}( \rm H\beta)$ plane was already proposed by \cite{Zackrisson_13} to identify high \fesc sources, predicting high \fesc to correlate with steep \uvbeta slopes and EW$_{0}( \rm H\beta)$<150 \AA{}. However, this is not seen in the LzLCS sample \citep{Flury_22b}. We indeed find a correlation between steep \uvbeta slopes and high \fesc:  below \uvbeta= -2.5 we find no low \fesc galaxies. This correlation stems from the fact that the addition of nebular continuum reddens the slope. For a young stellar population this reddening pushes the \uvbeta slope from $\sim$-3.0 up to $\sim$-2.5, almost independently of the chosen IMF \citep{Katz_24, Yanagisawa_24}. In the case of BPASS, it is only possible to have \uvbeta< -2.5 if the nebular continuum is reduced, due to a non-zero \fesc.
We do not see a relation with EW$_{0}( \rm H\beta)$. Many high \fesc galaxies have an EW<200 \AA{} but no clear trends with decreasing EW are visible. In the same figure we also highlight our high confidence sample of 71 galaxies with Bayes factor > 100 in the diamond markers. As expected, all of these sources have very blue \uvbeta slopes, mostly below -2.5, making the \fesc= 0 solution unlikely. 

\subsection{$f_{\rm esc}$ vs \uvbeta}
\begin{figure}
    \centering
    \includegraphics[width=\columnwidth]{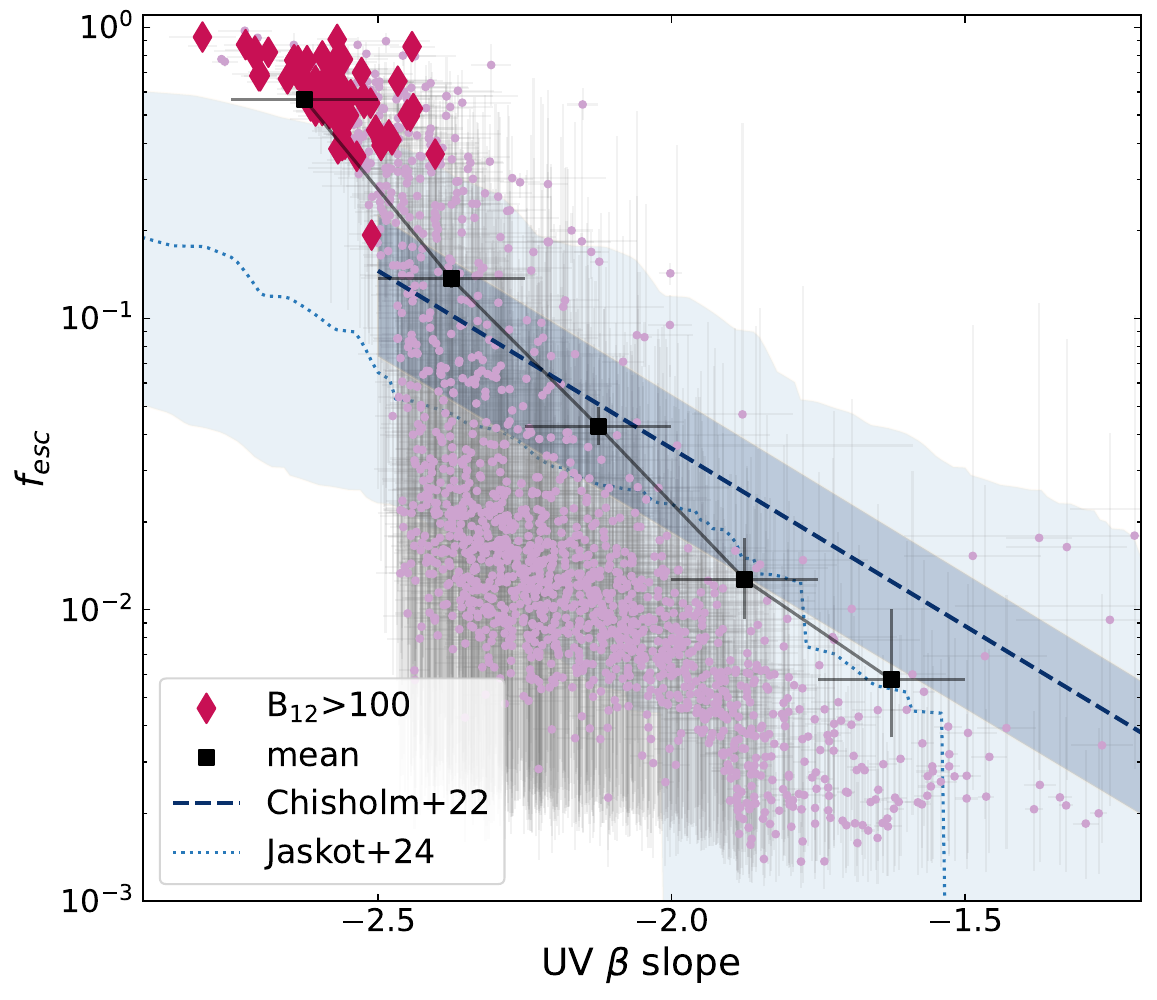}
    \caption{\fesc vs \uvbeta slope. The darker diamond points highlight the high confidence sample, while the dots represent the parent sample. The squares represent the average \fesc in \uvbeta bins, calculated from the parent sample. The \fesc-\uvbeta relation from \protect\cite{Chisholm_22} is shown as the dashed line, with the uncertainty shown as the shaded area. With the dotted line we show the results of \cite{Jaskot_24a}, using the \texttt{LyCsurv} code \citep{LyCsurv}. The average \fesc decreases as the \uvbeta slope becomes bluer.}
    \label{fig:avg_fesc_beta}
\end{figure}

In Figure~\ref{fig:avg_fesc_beta} we analyze the \fesc distribution as a function of the \uvbeta slope and the evolution of mean \fesc with bins of \uvbeta slope.
The mean \fesc is calculated over the whole parent sample. We find a consistent decrease of mean \fesc as slopes get shallower indicating an overall redder spectrum, as expected from our model. This trend is qualitatively consistent with the results of both \citet{Chisholm_22} and \citet{Jaskot_24a} shown in the same plot, although our relation is steeper. Both relations from the literature are based on a sample of 89 $z=0.3$ galaxies, with the main difference between their methods being the use of survival analysis in the case of \citet{Jaskot_24a}, using the \texttt{LyCsurv} code \citep{LyCsurv}. This indicates that on average galaxies that leak a lot of their ionizing continuum are those that have a steep \uvbeta slope. 

Here again we see the limit of \uvbeta$\sim$-2.5, as calculated from the best fit model, below which it is very difficult to have \fesc = 0, due to the impact of the nebular continuum on the slope. Although the overall shape of the relations is similar, it must be noted that the $\beta_{\rm UV}<-2.5$ range is almost unpopulated in the LzLCS sample, possibly due to more dust being present at low redshift.

We also see a large amount of scatter in the relation, which is not symmetric, toward lower values of \fesc. We find a larger scatter in the relation than that found by \cite{Chisholm_22}. Overall, the trend that we find seems to be mostly driven by the necessarily higher \fesc at low (\uvbeta\ <-2.5) \uvbeta values rather than by a real smooth decrease in \fesc with a reddened \uvbeta. The mean \fesc values indeed decrease exponentially to redder \uvbeta.
Around \uvbeta$\sim$-2.5 we see a steep drop in many \fesc values, which is likely due to the fact that the spectra of these sources do not contain significant information on the \fesc values such that they end up spanning almost the full range of the prior distribution. This also leads to a particularly large dispersion in \fesc values around \uvbeta of -2.5 to -2.2.

\subsection{\fesc vs M$_{\rm UV}$}

\begin{figure}
    \centering
    \includegraphics[width=\columnwidth]{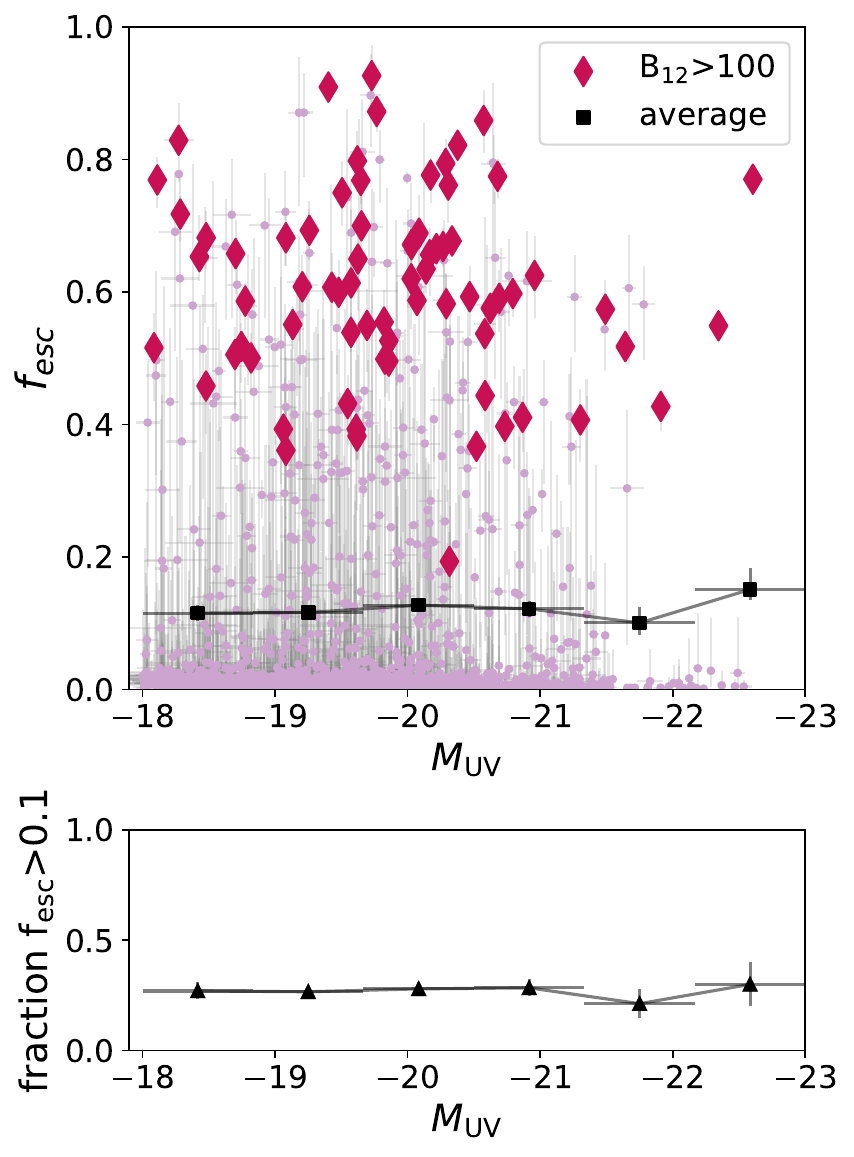}
    \caption{\textbf{Top:} \fesc vs $\rm M_{\rm UV}$. The parent sample is shown with the pink dots, the high confidence sample is the dark diamonds and the average \fesc in bins of $\rm M_{\rm UV}$ is shown as the black squares. The mean \fesc does not show a trend with $\rm M_{\rm UV}$. The average \fesc of our sample is consistently measured between 10-15\% in all bins.
    \textbf{Bottom:} The fraction of sources with $f_{\rm esc}$>0.1 in each UV magnitude bin. This fraction also shows no trend with $\rm M_{\rm UV}$.}
    \label{fig:fesc_MUV}
\end{figure}

To fully understand reionization it is important to identify which subpopulation of galaxies leaks the most ionizing photons, faint or bright ones. For this reason, we show the relation between \fesc and M$_{\rm UV}$ in Figure~\ref{fig:fesc_MUV}. 

Both the high confidence sample and the average \fesc do not show a clear trend with M$_{\rm UV}$. At all magnitudes, the average \fesc sits at around 10\%, consistent with the results of \cite{Mascia_25}, but in contrast with the results of \cite{Papovich_25} who find much lower average \fesc, $\langle  f_{\rm esc} \rangle \sim 3\%$.

In the bottom panel of the same figure, we present the fraction of objects with \fesc> 0.1 in each M$_{\rm UV}$ bin. Also this remains mostly flat with M$_{\rm UV}$. 

This implies that the trends in mean \fesc are due to the amount of leaker galaxies compared to non-leakers. 
Tabulated values of both the average \fesc and the fraction of galaxies with \fesc > 0.1 in each magnitude bin can be found in Table~\ref{tab:fesc_MUV} for ease of reading. 

Our values are consistent with current models of reionization, which suggest that mean \fesc values of about 5\%-20\% are needed to reionize the universe by $z \sim 6$  \citep{Bouwens_15, Robertson_15, Finkelstein_19}. Our results would thus imply that bright and faint galaxies have an equal number of strong leakers and that both bright and faint galaxies contribute. A more detailed discussion will be presented in Giovinazzo et al, in prep. 
We note, however, that it is possible that our sample is lacking some high \fesc sources at faint magnitudes, as faint sources with no lines might not have high-confidence redshifts in the DJA or might not have been targeted. It is therefore possible that our inferred average \fesc at faint magnitudes could be somewhat underestimated.

\begin{table}
    \centering
    \renewcommand{\arraystretch}{1.2} 
    \caption{Tabulated values of mean and standard deviation of the $\langle f_{\rm esc} \rangle$ distribution with $\rm M_{\rm UV}$ (top panel of Figure~\ref{fig:fesc_MUV}) and fraction of sources with $f_{\rm esc}$>0.1 (bottom panel of Figure~\ref{fig:fesc_MUV})}
    \label{tab:fesc_MUV}
    \renewcommand{\arraystretch}{1.5} 

    \begin{tabular}{ccc}
    \hline
    $M_{\rm UV}$ & $\langle \rm f_{\rm esc} \rangle$ & fraction(\fesc>0.1) \\

\hline
 -18.42 & 0.12$_{-0.01}^{+0.01}$ &      0.27  \\
 -19.25   & 0.12$_{-0.01}^{+0.01}$ &      0.27 \\
 -20.08 & 0.13$_{-0.01}^{+0.01}$ &      0.28 \\
 -20.92 & 0.12$_{-0.01}^{+0.01}$ &      0.28 \\
 -21.75   & 0.10$_{-0.02}^{+0.02}$ &      0.23 \\
 -22.58 & 0.15$_{-0.02}^{+0.03}$ &      0.30   \\

\hline
\end{tabular}

\end{table}

\subsection{The cumulative distribution function of \fesc}
\begin{figure}
    \centering
    \includegraphics[width=\columnwidth]{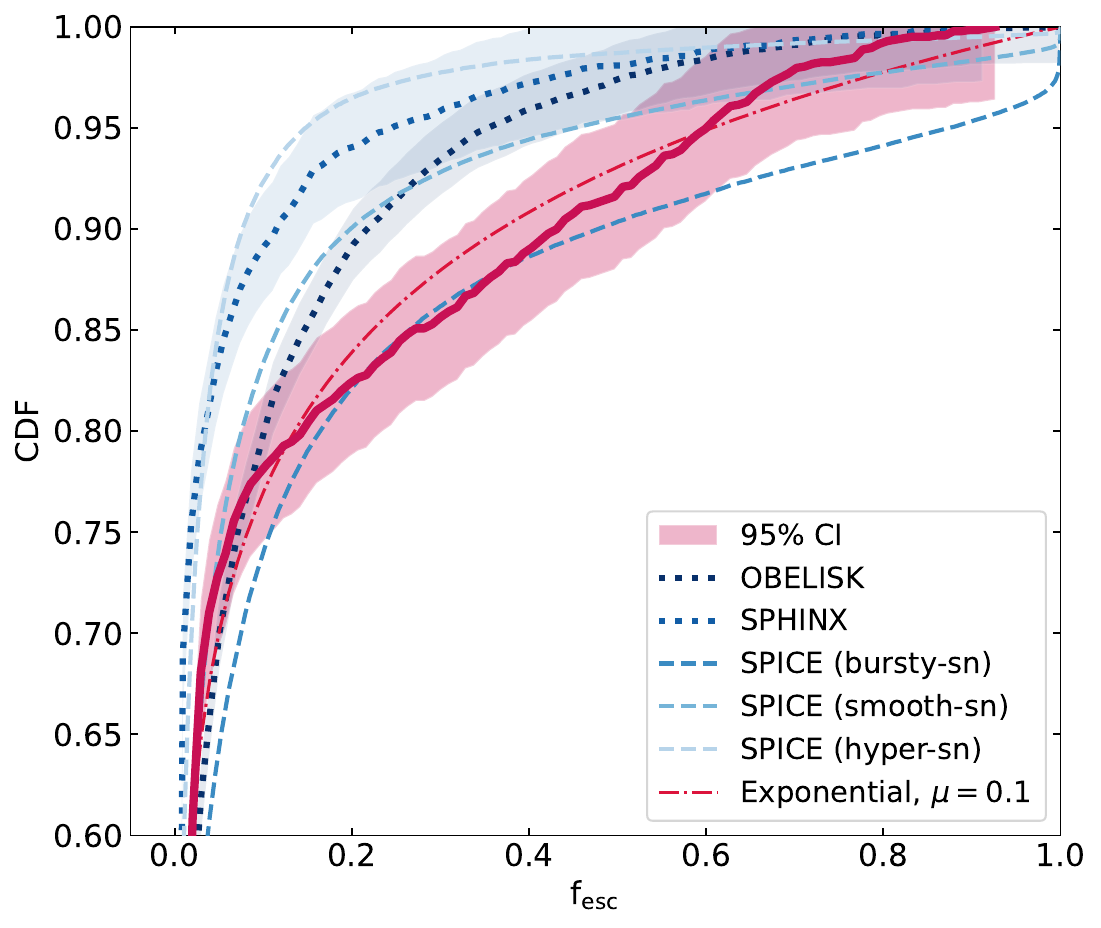}
    \caption{Empirical cumulative distribution function (CDF) of \fesc in the parent sample. The dark pink solid line is the CDF for our data, with the shaded area representing the 95\% confidence interval. The dotted lines with the shaded areas show the CDF of different simulations, including \textsc{Sphinx}  and  \textsc{Obelisk} with the respective 95\% confidence interval. The dashed blue lines show the three realizations of the \textsc{Spice} simulation. The dot-dashed purple line is a simple exponential distribution with a mean of 10\%, as justified by the results of Figure~\ref{fig:fesc_MUV}, and which fits our CDF rather well. All models consistently predict mostly low \fesc objects with a few high \fesc sources that drive the mean to non-negligible values. }
    \label{fig:CDF}
\end{figure}

A good test to determine if our \fesc distribution is consistent with reionization models is to compare the cumulative distribution function (CDF) of our inferred \fesc values with simulations that also estimate \fesc. We compare our CDF to the CDF of three simulations: (1) \textsc{Sphinx} \citep{Rosdahl_22, Katz_23}, which we calculate from the entire SPDRv1, (2) \textsc{Obelisk} \citep{Trebitsch_21}, which we calculate from all the main snapshots at $5<z<9.6$, and (3) the three models of SPICE \citep{Bhagwat_24}. For both \textsc{Sphinx} and \textsc{Obleisk} we apply the same M$_{\rm UV}<-18$ cut as we did for the parent sample. Although we apply the same cuts to the simulations as we do to the observations, it is important to note that the selection function of the observed sample is not well characterized, making the comparison between observations and simulations not obvious. Our results are shown in Figure~\ref{fig:CDF}. 

We also show the CDF of an exponential distribution of \fesc, similar to that presented in \citet{Kreilgaard_24} but with a mean \fesc of 10\%, which is more in line with our measurements, as shown in Figure~\ref{fig:fesc_MUV}.
Our CDF shows that most of the sample has extremely low \fesc and few galaxies have very high \fesc, with only about 20\% of the sample showing \fesc > 0.2. This is broadly consistent with all of the simulations, whose CDFs all show a large number of very low \fesc galaxies, but differ in the amounts of high \fesc sources. Both \textsc{Obelisk} and \textsc{Sphinx} contain a much larger fraction of sources with very low \fesc with respect to our sample, which could be due to ionizing photons produced by obscured star formation, which are not taken into account in our model but are in the simulations. Taking this source of ionizing photons into account can lower the inferred escape fraction possibly leading to the discrepancy between \textsc{Obelisk} and \textsc{Sphinx} and our model.

The simulation that is most consistent with our results is the \textit{bursty-sn} model of SPICE. Out of all the SPICE models, this is the only one which reionizes at a time consistent with observations \citep{Bhagwat_24}, which is very encouraging for our results. The exponential model is also entirely within our confidence interval, showing good agreement with our data. This indicates that our method results in a consistent fraction of high \fesc galaxies, compatible with current models of reionization, even if we cannot well determine the completeness of our sample.

\subsection{High confidence vs parent sample}

\begin{figure}
    \centering
    \includegraphics[width=\columnwidth]{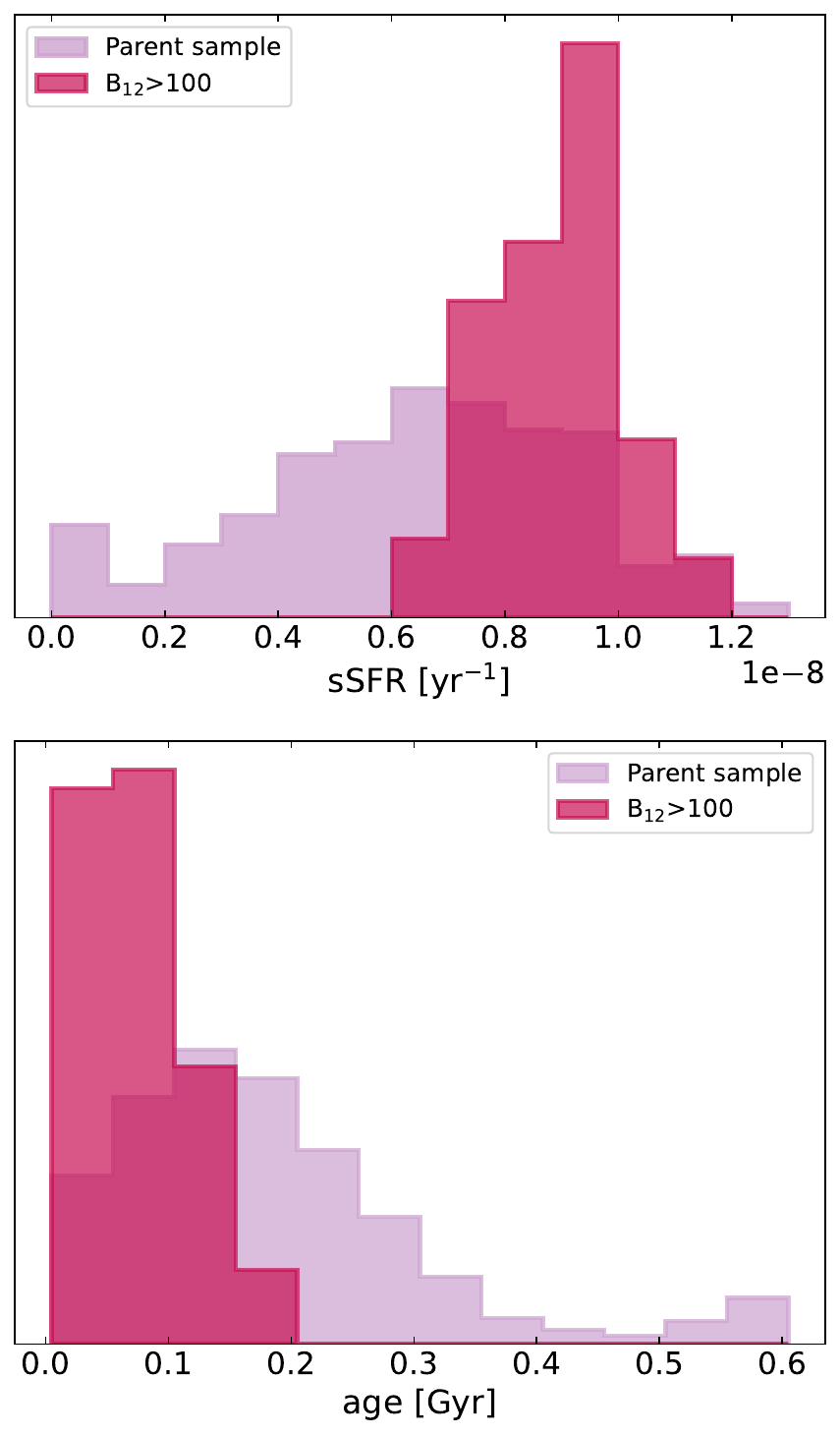}
    \caption{\textbf{Top: }Comparison of the specific star formation rate distribution between the parent sample (light pink) and the high confidence sample(dark pink). \textbf{Bottom: }Comparison of the age distribution between the parent sample and the high confidence sample. The high confidence sample is both more star forming and younger, indicating a relation between these properties and Lyman continuum leakage.}
    \label{fig:sSFR_age}
\end{figure}

Next, we want to test if the high-confidence, high \fesc sample has any particular properties that sets it apart from the parent population at the same redshift. To do this, we  compare the physical properties derived from \texttt{bagpipes} between our high confidence sample and the parent sample. We present this comparison in Figure~\ref{fig:sSFR_age}. 
Looking at the specific star formation rate (sSFR = SFR$_{10}$/M$_{*}$) distributions, it is clear that on average the high confidence sample has higher sSFR than the parent sample, in line with the model of \cite{Ferrara_25}. Our candidate leakers are thus likely to have gone through a recent burst of star formation, consistent with results from simulations \citep[e.g.][]{Trebitsch_17}. This is also seen in the age distributions, which indicate that all the high-confidence \fesc sources have ages <0.2 Gyr while the parent sample has sources with ages up to 0.6 Gyr.

From this figure we can say that overall our sample of confident leakers is more star forming and younger than the parent sample. This is consistent with the conditions for production and escape of ionizing photons as young stars are those that can produce Lyman continuum photons and are also required to produce very steep UV slopes (\uvbeta< -2.5).

\section{Comparison with other methods to indirectly estimate \fesc}\label{ch:Comparison}

There have been many attempts in the literature at linking \fesc to other characteristics of galaxies, such as the [\ion{O}{III}]$\lambda5007$/[\ion{O}{II}]$\lambda3727$ line ratio \citep[$O_{32}$; e.g.,][]{Jaskot_13,Nakajima_Ouchi_14, Izotov_18b, Paalvast_18, Tang_21}, the \uvbeta slope \citep[e.g.,][]{Chisholm_22, Flury_22b}, star formation rate surface density \citep[$\Sigma_{\rm SFR}$;][]{Naidu_20}, and others that have been already discussed in Section~\ref{ch:introduction}.
In this section, we will explore some of these methods and determine if our data follows the relations previously found in literature. 

\subsection{The O32 line ratio}

\begin{figure}
    \centering
    \includegraphics[width=\columnwidth]{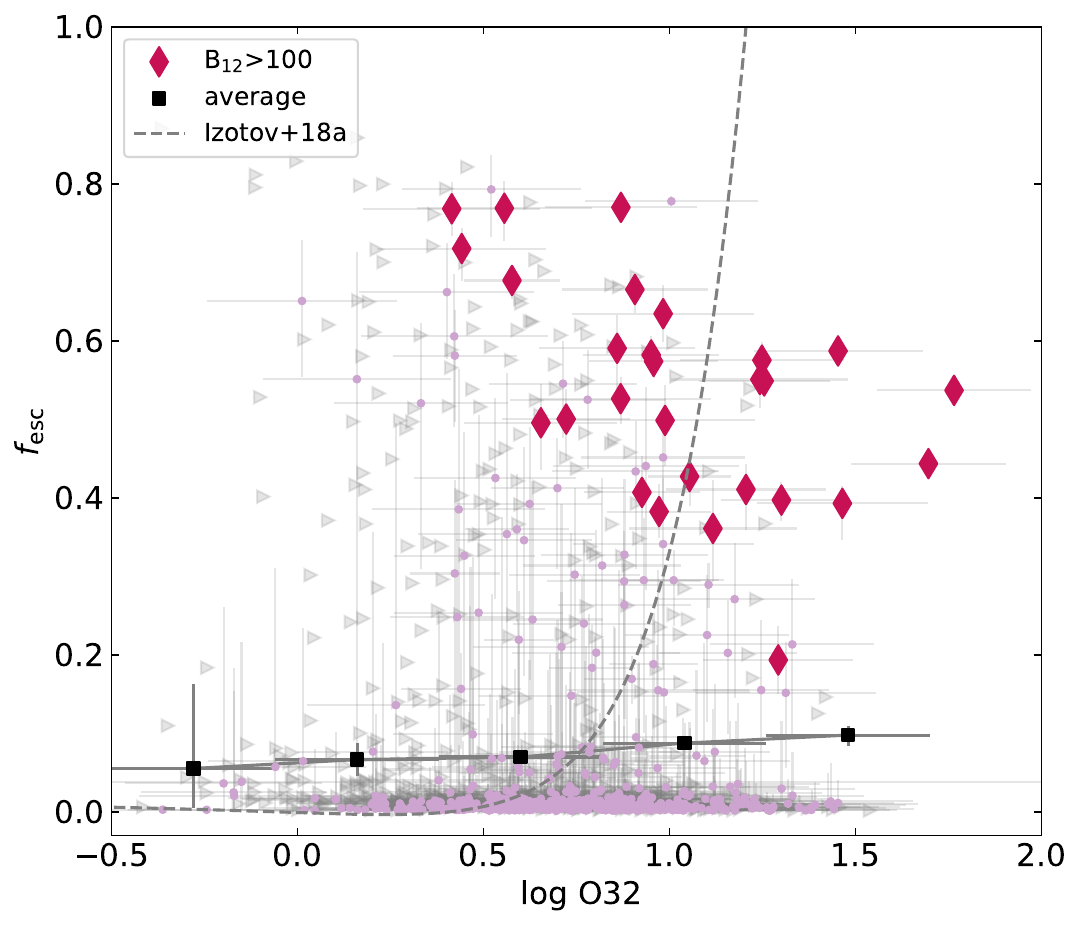}
    \caption{Relation between \fesc and O$_{32}$, with the average \fesc as black squares, in bins on log(O$_{32}$). The diamond points are the high confidence sample and the dots are the parent sample. The grey dashed line is the relation between \fesc and O$_{32}$ form \citet{Izotov_18a}. The grey triangles indicate lower limits on the O$_{32}$ ratio for the points with SNR([\ion{O}{II}])<3. Our points show no strong trend between $\langle$\fesc$\rangle$ and O$_{32}$.}
    \label{fig:O32}
\end{figure}

We first explore O$_{32}$, defined as [O~{\sc iii}]$\lambda$5007/[O~{\sc ii}]$\lambda$3727, an indicator for \fesc which has been studied both from simulations and observations, but for which a clear understanding of its suitability to estimate \fesc is still lacking. High O$_{32}$ ratios might be indicative of density-bounded conditions and therefore high \fesc, but its dependence on ionization parameter and metallicity \citep{Sawant_21} makes its interpretation challenging. Simple photoionization models show a correlation between O$_{32}$ and low optical depth for LyC photons \citep{Jaskot_Oey_13, Nakajima_Ouchi_14} while results from simulations show no clear trends between \fesc and O$_{32}$ \citep{Katz_20,Kirk_20,Choustikov_24}. Low redshift observations also show a tenuous relation with varying degree of scatter in the \fesc - O$_{32}$ plane \citep{Faisst_16, Nakajima_20, Flury_22b}, overall showing that while many LyC leakers show a high O$_{32}$, this condition alone is not enough to identify leakers. This is also seen in our sample in Figure~\ref{fig:O32}. Many sources with high O$_{32}$ ratios do indeed show \fesc>0.2, but the average values in bins of O$_{32}$ only increase very slowly. This result is consistent with those of \cite{Choustikov_24} who also find no clear trend between \fesc and O$_{32}$ in the \textsc{Sphinx} simulation. 

It is important to note here that the relation between O$_{32}$ and \fesc was proposed for a density bounded scenario, whereas we assume a picket fence model in our SED fitting, which could be the cause of the discrepancy with the low-redshift  observations. However, the agreement with the result of \textsc{Sphinx} by \citet{Choustikov_24} could imply that this relation might not be easily applicable at high redshift.

\subsection{A high SFR surface density}

\begin{figure}
    \centering
    \includegraphics[width=\columnwidth]{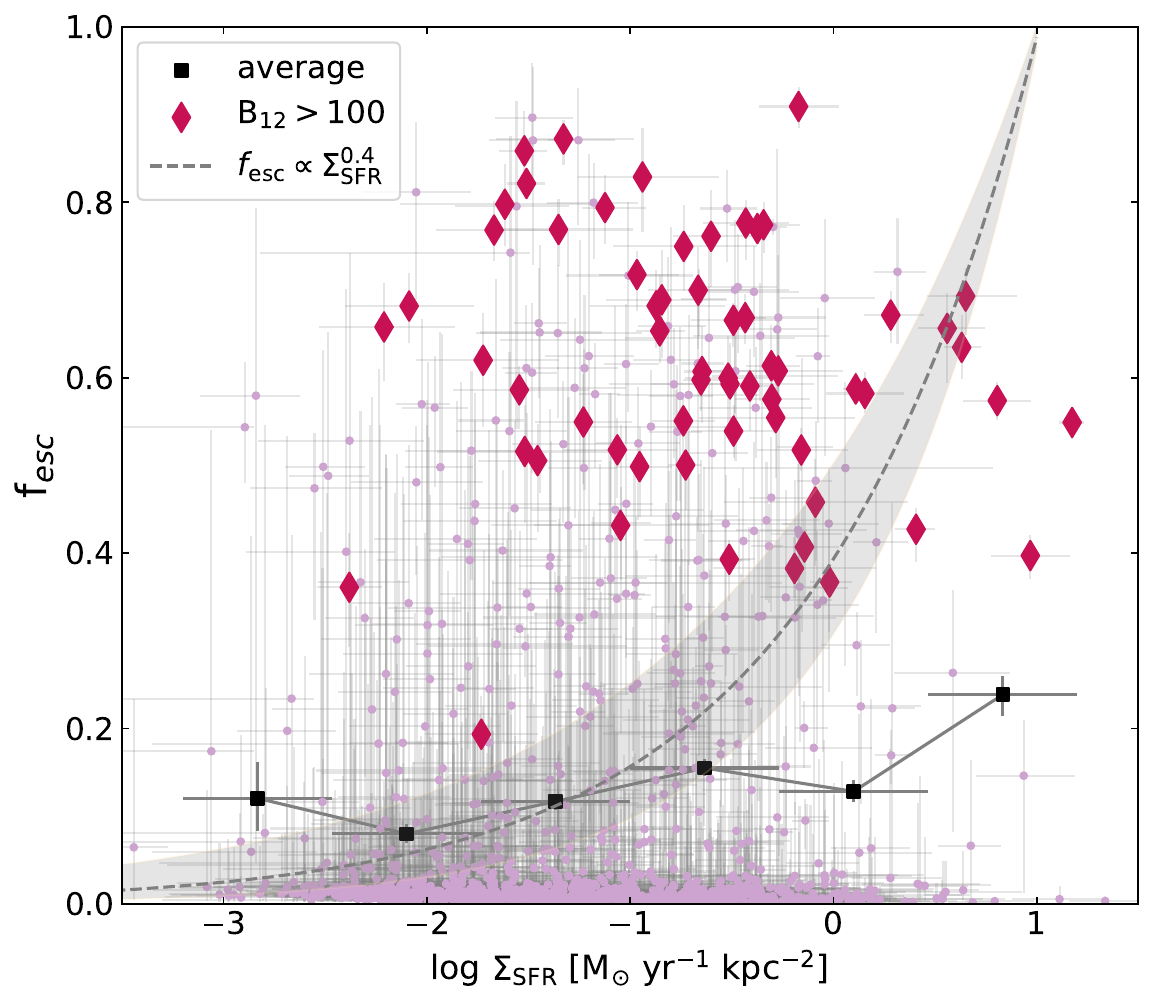}
    \caption{$f_{\rm esc}$ vs star formation rate surface density. The high confidence sample is shown with the dark diamonds, the average $f_{\rm esc}$ is shown with the black squares.
    Our sample does not show any significant trend of higher \fesc with higher $\Sigma_{\rm SFR}$, as was expected from the proposed relation from \protect\cite{Naidu_20} that is shown as the dashed line.}
    \label{fig:fesc_simgasfr}
\end{figure}

We also look at the star formation rate surface density, defined as
\begin{equation}
    \Sigma_{\rm SFR} = \frac{\rm SFR/2}{\pi \rm r_{\rm eff}^2}
\end{equation}
where we use SFR$_{10}$ for the star formation rate.
Lyman continuum leakers have previously been found to be generally compact with high $ \Sigma_{\rm SFR}$ \citep{Sharma_16, Sharma_17,Naidu_20, Flury_22b}. Others found that UV compact sources are more likely to be LyC leakers \citep{Marchi_18}. This suggests that highly concentrated star-forming regions might create the ideal conditions for feedback to clear paths in the ISM to leak ionizing radiation \citep{Flury_22b}. In Figure~\ref{fig:fesc_simgasfr} we show \fesc vs $ \Sigma_{\rm SFR}$ for our sample. 

We compare this to the proposed relation from  \cite{Naidu_20}, \fesc$\propto (\Sigma_{\rm SFR})^{0.4}$. Our sample clearly does not follow this simple scaling. Although there may be a mild trend to higher escape fractions, on average, at higher $\Sigma_{\rm SFR}$, this is not significant. There is again enormous scatter in the individual measurements, and it is certainly not the case that all sources with very high SFR surface densities have high escape fractions. 

In summary, we conclude that none of the previously proposed, `simple' relations to estimate \fesc are consistent with our results, apart from the strong trend with UV continuum slope.

\section{Discussion}\label{ch:discussion}

\subsection{Selection}

Our choice to use all the NIRSpec/prism spectra from the entire DJA gives us access to an unprecedented amount of deep rest-optical spectra in the redshift range $5<z<10$, but it also comes with important considerations about the selection. First of all, the choice of using only sources with grade=3, although necessary to ensure that the objects are fit at the correct redshift, biases the sample towards line emitters. As the grades are assigned through visual inspection, sources with stronger Lyman breaks and stronger lines are more likely to get a high grade. Because of this, we are possibly missing leakers with very weak lines, especially at the faint end of our magnitude range. 

Second, the choice of using a wide array of different programs makes the selection function essentially intractable. Many of the spectra used here were only included as filler targets in individual masks. It is therefore difficult to know if our results are valid for a UV- or mass-selected sample of galaxies, as it is unclear what kind of biases enter our sample through the mask design strategies of the different programs. This will have to be tested with future initiatives that aim to obtain mass-selected galaxy samples with prism spectroscopy.

\subsection{Method caveats}

Until now, we have focused on the role of the \uvbeta slope and the Balmer lines as indicators of high \fesc that can be used in the EoR. However, there is another possible scenario that could lead to a source having a steep \uvbeta but weak Balmer lines, which is the case of a recently quenched source that has had star formation within the last 100 Myr \citep{Looser_24}.
In such cases, O-type stars with lifetimes of $\lesssim 10$ Myr have already died (reducing the observed nebular emission), while less massive B-type stars (with lifetimes of $\sim$100 Myr) would still be present, producing slightly steeper UV slopes.
This indicates a possible degeneracy of our results with the SFH, as most of our high-confidence sources have a very blue \uvbeta slope but weak H$\beta$. This is exemplified in the third panel in Figure~\ref{fig:rubies_spec}, where we compare the SFH of a single galaxy for the high \fesc case and for the no \fesc case. However, the same example also shows that even with a recent quenching SFH, the \fesc = 0 cannot quite match the data as well, as quantified by the very high Bayes factor. It is important to note that we fit everything together, \fesc and SFH, and therefore this degeneracy is included in the uncertainties of \fesc, which are small for our high-confidence sample. We also note that none of these high-confidence sources shows any signs of a Balmer break, which one could expect to appear if the star-formation was suppressed for long enough.
Nevertheless, it will be important to fit sources with known, directly measured escape fractions \textit{and} with NIRSpec/prism spectra to further test our method in the future.

\section{Summary}\label{ch:Conclusion}

In this work, we present the addition of a picket-fence model to \texttt{bagpipes} as a new tool to recover \fesc from the spectra of EoR galaxies. This allows us to estimate this important parameter for reionization directly, without the need for indirect tracers calibrated at lower redshift.

Using public NIRSpec/prism spectra and photometry available in the DJA, we compile 1'428 galaxies at redshifts $5<z_{\rm spec}<10$ from a variety of public programs (see Table~\ref{tab:DJA}).
We derive the physical properties and \fesc of each source through SED fitting on the spectra, using the picket-fence model with \texttt{bagpipes}. We perform a second \texttt{bagpipes} run with \fesc = 0 and quantify the best fit model with the use of the Bayes factor. Doing this, we identify 71 galaxies as high-confidence Lyman continuum emitters, based on their Bayes factor ($B_{12}>100$).

We validate our picket fence model by performing a recovery simulation with various noise levels and find that the SED fitting can recover the correct \fesc, with the exception of low \fesc values at low SNR, which are underestimated. High \fesc values are always recovered correctly with high confidence ($B_{12}>100$). 

The main results from our analysis are:
\begin{itemize}
    \item The high-confidence LyC leaker sample is mostly characterized by blue \uvbeta slopes, $<-2.5$. The average \fesc of the overall sample decreases with increasing \uvbeta slope, broadly consistent with results from \cite{Chisholm_22} and \cite{Jaskot_24a}. This shows that the \uvbeta is an important indicator for \fesc also at high redshift, when applied as a general trend. 
    \item There is no significant trend of \fesc as a function of M$_{UV}$.
    Overall we find average \fesc values at all magnitudes hovering around 10-15\%. This is consistent with the required values for completing reionization  with galaxies alone by $z\sim 6$, as discussed in the past literature. 
    \item The cumulative distribution function of the \fesc measured with our method is consistent with simulations. We find that most of our sources have very low \fesc with only a few high \fesc sources, as expected. The closest model to our CDF is the \texttt{bursty-sn} model of SPICE \citep{Bhagwat_24}. We find our CDF to be consistent with an exponential distribution of \fesc with a mean of 10\%.
    \item We find that the high-confidence sample is on average younger and more star forming than the overall sample, indicating that both properties might aid the escape of ionizing photons. 
    \item We do not find any trend between \fesc and O$_{32}$. This and the overall shape of the distribution is consistent with results from the \textsc{Sphinx} simulation \citep{Choustikov_24}. 
    We also do not find a significant trend between \fesc and $\Sigma_{\rm SFR}$. 
\end{itemize}

Overall, we show that our method produces results consistent with our theories of reionization and with previous studies. This suggests that it is indeed possible to estimate \fesc directly in the epoch of reionization, thanks to the capabilities of the \textit{JWST}, with a caveat being the degeneracy with SFH, which could eventually be broken through the use of high resolution spectroscopy in the UV to search for stellar absorption lines. 

With \fesc being the key unknown of reionization, the ability to get a consistent estimate of its value for galaxies at high redshift will allow us to head towards a better understanding of the epoch of reionization, as more and more data are acquired by \textit{JWST}. Future work will necessarily have to include better estimates of the ionizing photon production efficiency in high redshift galaxies, possibly including a model to directly calculate the rate of LyC photon productions in galaxies (Giovinazzo et al., in prep). 

\begin{acknowledgements}
    
This work is based on observations made with the NASA/ESA/CSA James Webb Space Telescope. The raw data were obtained from the Mikulski Archive for Space Telescopes at the Space Telescope Science Institute, which is operated by the Association of Universities for Research in Astronomy, Inc., under NASA contract NAS 5-03127 for \textit{JWST}. 
Some of the data products presented herein were retrieved from the Dawn JWST Archive (DJA). DJA is an initiative of the Cosmic Dawn Center, which is funded by the Danish National Research Foundation under grant No. 140 (DNRF140).
      
This work has received funding from the Swiss State Secretariat for Education, Research and Innovation (SERI) under contract number MB22.00072, as well as from the Swiss National Science Foundation (SNSF) through project grant 200020\_207349.
\end{acknowledgements}




\bibliographystyle{aa}
\bibliography{bib} 




\appendix



\onecolumn

\section{High-confidence sample}

\begin{figure}[h]
    \centering
    \includegraphics[width=\linewidth]{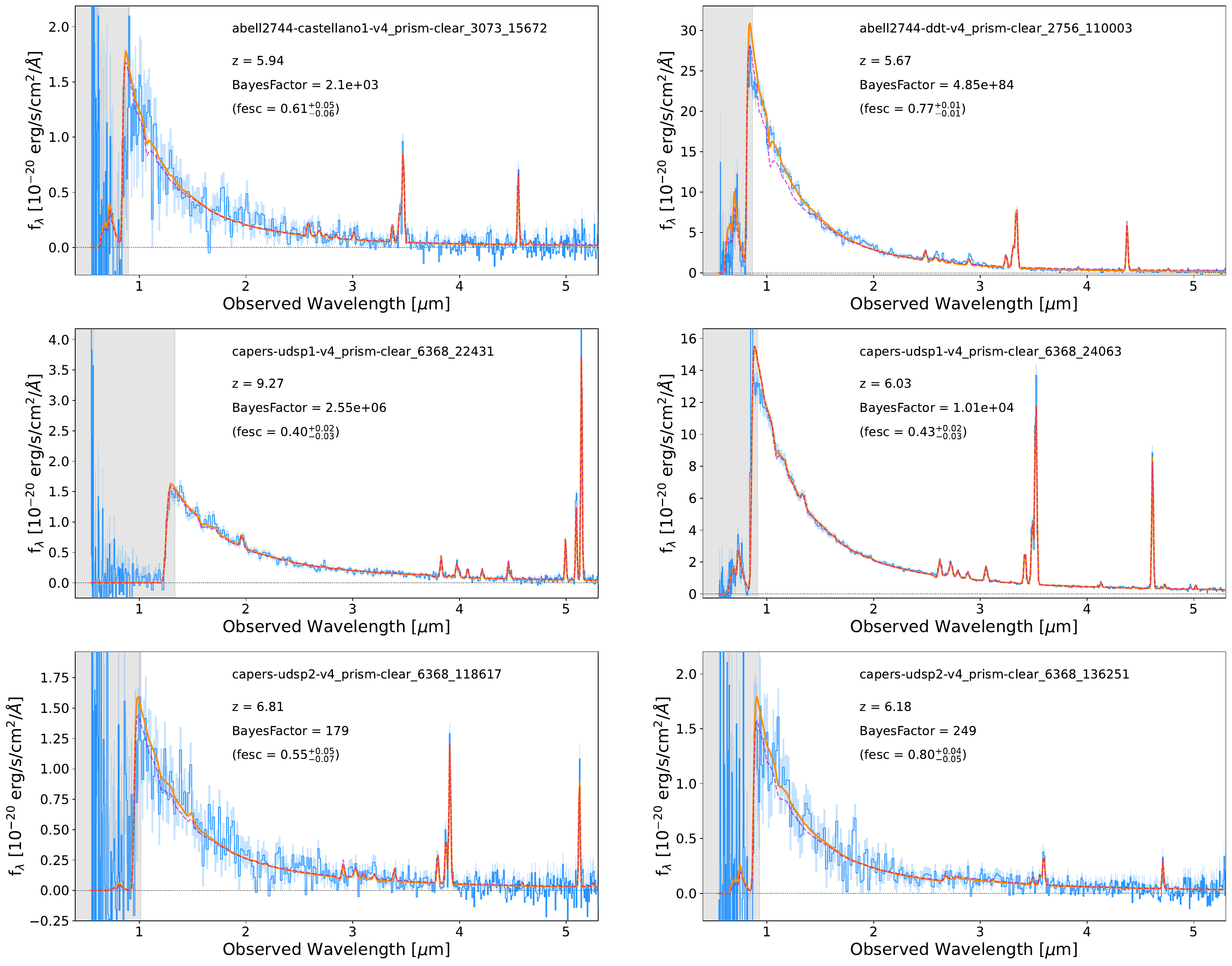}
    \caption{Fits of high confidence sample sources. Just as in Figure ~\ref{fig:rubies_spec} The orange line represents the high \fesc fit, the magenta dashed line represents the $f_{\rm esc} = 0$. The grey shaded area represents the region masked when performing the fit.}
    \label{fig:appendix}
\end{figure}

Figure~\ref{fig:appendix} shows the SED fits to the spectra of some of the 71 galaxies in the high-confidence LyC emitter sample with $B_{12}>100$ and Table~\ref{tab:appendix} lists their properties.

\begin{sidewaystable*}
        \caption{Properties of the high  confidence sample. (The full table will be available online.)}
    \label{tab:appendix}
    \centering
\renewcommand{\arraystretch}{1.5} 

\begin{tabular}{lccccccccc}
\hline
 Source                                            &    $z_\mathrm{spec}$ &     RA &    Dec &   \fesc &   logM [$M_\odot$] &    logSFR$_{10}$ [$M_{\odot} \rm yr^{-1}$] &   $\rm M_{\rm UV}$  &   \uvbeta &  EW(H$\beta$) \\
\hline
 abell2744-castellano1-v4\_prism-clear\_3073\_15672 & 5.94 &   3.51380  & -30.35334  &   $0.61_{-0.06}^{+0.05}$ &   $7.56_{-0.14}^{+0.21 }$ &    $-0.48_{-0.12}^{+0.14}$ & $-19.43_{-0.06}^{+0.06}$ &  $-2.58_{-0.05}^{+0.06}$ & $105_{-122}^{+122}$  \\
 abell2744-ddt-v4\_prism-clear\_2756\_110003        & 5.67 &   3.59069 & -30.39554  &   $0.77_{-0.01}^{+0.01}$ &   $8.50_{-0.03}^{+0.06}$ &     $0.49_{-0.02}^{+0.03}$ & $-22.60_{-0.01}^{+0.01}$ &  $-2.64_{-0.01}^{+0.01}$ &  $83_{-13}^{+13}$ \\
capers-udsp1-v4\_prism-clear\_6368\_22431          & 9.27 &  34.46025  &  -5.18500   &   $0.40_{-0.03}^{+0.02}$ &   $7.92_{-0.04}^{+0.06}$ &    $-0.08_{-0.03}^{+0.05}$ & $-20.73_{-0.02}^{+0.02}$ &  $-2.56_{-0.02}^{+0.02}$ & $199_{-89}^{+89}$ \\
capers-udsp1-v4\_prism-clear\_6368\_24063          & 6.03 &  34.50378 &  -5.19384 &   $0.43_{-0.04}^{+0.02}$ &   $8.91_{-0.03}^{+0.03}$ &     $0.95_{-0.03}^{+0.03}$ & $-21.91_{-0.01}^{+0.01}$ &  $-2.48_{-0.01}^{+0.01}$ &  $67_{-4}^{+4}$ \\
capers-udsp2-v4\_prism-clear\_6368\_118617         & 6.81 &  34.47512  &  -5.11907 &   $0.55_{-0.07}^{+0.05}$ &   $8.01_{-0.15}^{+0.16}$ &    $-0.06_{-0.15}^{+0.13}$ & $-19.82_{-0.05}^{+0.05}$ &  $-2.59_{-0.05}^{+0.05}$  & $153_{-115}^{+115}$ \\
... & ... & ... & ... & ... & ... & ... & ... &... & ... \\ 
\end{tabular}

\end{sidewaystable*}


\end{document}